\documentclass[aps,prb,groupedaddress,superscriptaddress,reprint]{revtex4-1}
\usepackage{graphicx}
\usepackage{dcolumn}
\usepackage{amsthm}
\usepackage{bm}
\newcommand{\ket}[1]{|#1\rangle}
\newcommand{\bra}[1]{\langle#1|}

\newcommand{\sign}{\text{sgn}}
\newcommand{\trace}{\mathop{\mathrm{tr}}}

\pdfoutput=1
\usepackage{graphicx,graphics,epsfig,subfigure,times,bm,bbm,amssymb,amsmath,amsthm,mathrsfs,MnSymbol}
\usepackage{gensymb}
\usepackage{amsfonts}
\usepackage[matrix,frame,arrow]{xypic}
\usepackage[pdfstartview=FitH]{hyperref}
\hypersetup{
    colorlinks=true,       
    linkcolor=red,          
    citecolor=magenta,        
    filecolor=magenta,      
    urlcolor=cyan,           
    runcolor=cyan
}
\usepackage{epstopdf}
\usepackage[pdftex]{color}
\usepackage{hypernat}
\usepackage{braket}
\usepackage{enumerate}
\usepackage[normalem]{ulem}
\usepackage[usenames,dvipsnames]{xcolor}
\usepackage{multirow}
\usepackage{mathtools}

\definecolor{orange}{rgb}{1,0.5,0}

\newcommand{\ignore}[1]{}
\usepackage{geometry}

\begin{document}

\title{ Dynamical quantum phase transition for mixed states in open systems}

\author{Haifeng Lang}
\affiliation{Institute of
Physics, Chinese Academy of Sciences, Beijing 100190, China}
\affiliation{Kirchhoff-Institute for Physics, Heidelberg University, 69120 Heidelberg, Germany}

\author{Yixin Chen}
\affiliation{College of Chemistry and Molecular Engineering, Peking University, Beijing 100871, China}

\author{Qiantan Hong}
\affiliation{School of Physics, Peking University, Beijing 100871, China}

\author{Heng Fan}
\email{hfan@iphy.ac.cn}
\affiliation{Institute of
Physics, Chinese Academy of Sciences, Beijing 100190, China}
\affiliation{CAS Center for Excellence in Topological Quantum Computation, University of Chinese Academy of Sciences, Beijing 100190, China}





\begin{abstract}
Based on a kinematic approach in defining a geometric phase for a density matrix, we define the generalized Loschmidt overlap amplitude (GLOA) for an open system for arbitrary quantum evolution. The GLOA reduces to the Loschmidt overlap amplitude (LOA) with a modified dynamic phase for unitary evolution of a pure state, with the argument of the GLOA well-defined by the geometric phase, thus possessing similar physical interpretation to that of the LOA. The rate function for the GLOA exhibits non-analyticity at a critical time, which corresponds to the dynamical quantum phase transition. We observe that the dynamical quantum phase transition related to GLOA is not destroyed under a finite temperature and weak enough dissipation. In particular, we find that a new type of dynamical quantum phase transition emerges in a dissipation system. The proposed GLOA provides a powerful tool in the investigation of a dynamical quantum phase transition in an arbitrary quantum system, which not only can characterize the robustness of the dynamical quantum phase transition but also can be used to search for new transitions.
\end{abstract}

\pacs{Valid PACS appear here}
\maketitle


\section{Introduction.}
Nonequilibrium physics has a long history and has been attracting a great deal of interest. Recently, developments of experimental \cite{exper1,exper2,exper3,exper4,exper5,exper6,prlexpDQPT,natpexpDQPT,natexpDQPT}, numerical \cite{num1,num2,num3,num4,num5,num6} and theoretical \cite{theo1,theo2,theo3,theo4,theo5} techniques have shed light on nonequilibrium physics, especially issues induced by quenching, including the eigenvector thermalization hypothesis, equilibration after quenching and so on. For reviews of quenching in a closed pure state system, see Refs. [\onlinecite{PolkovnikovRMP,JensNatp}].
However, some problems remain open; for example, what are the dynamical and thermodynamical signatures following quenching in a mixed state? If natural unavoidable dissipations occur, how will they influence the properties of the system? Can the system exhibit some novel phenomena after engineered driven dissipation quenching?
Though there have been great achievements recently \cite{blabladiss,tempblabla,solvablediss}, we still know little about such topics.
This work is intended to answer these three questions from the viewpoint of a dynamical quantum phase transition (DQPT). We analyze the dynamical signature after quenching via GLOA, the generalization of the Loschimdt overlap amplitude.

LOA is a well-defined quantity in unitary evolution for a pure state. Its form is similar to the boundary partition function \cite{MHeyl},
\begin{eqnarray}
\mathcal{G}(t)=\bra{\psi_0}e^{-iH_ft}\ket{\psi_0}
\end{eqnarray}
where $ \ket{\psi_0} $ is the initial state, usually prepared as the ground state of the initial Hamiltonian $H_i$, which is equivalent to a boundary condition, and $e^{-iH_ft}$ is the evolution operator after quenching. We set $\hbar=1$ in this article for convenience. The non-analytical behavior of the rate function for LOA, $g_0(t)=-\frac{1}{L}\ln |\mathcal{G}(t)|^2$, where $L$ is the total degrees of freedom of the systems, is the finger-print of DQPT.
Time plays the same role in DQPT as temperature in an equilibrium phase transition, suggesting that DQPT can be referred to as a phase transition in the time domain.
The occurrence of DQPT provides the possibility of a quantum phase transition for the Hamiltonian before and after quenching \cite{MHeyl,RevDQPT}, although many counter examples have been reported \cite{vio1,vio2}. DQPT has been recently observed in both ultracold atoms and trapped-ion systems \cite{prlexpDQPT,natpexpDQPT,natexpDQPT}.
It is expected that DQPT is robust, i.e., it can occur when the initial state is mixed and the system is influenced by the environment.
However, for an open quantum system with a mixed state, a well-accepted generalized LOA showing the DQPT under extensive circumstances is still absent.

\section{Loschmidt overlap amplitude and dynamical topological order parameter.}
The mathematical structure of an equilibrium phase transition and partition function, for instance, Fisher zero points \cite{MHeyl}, universality class and renormalization group \cite{universalityclass} can be generalized to DQPT and LOA. In addition to the mathematical beauty, LOA has physical meanings, i.e., the Fourier transform of the power distribution up to a phase factor. In addition, the square of the modular of LOA equals the survival probability \cite{MHeyl}. Furthermore, topological information for the evolution is encoded in the argument of LOA, which is the total phase during evolution \cite{nu_Dprb}.

It is possible to construct a dynamical topological order parameter (DTOP) for DQPT \cite{nu_Dprb}. A varying DTOP reflects the occurrence of DQPT. The geometric phase, which can be easily obtained from the argument of LOA, is crucial for DTOP. When the system is noninteracting and has translation-invariance symmetry, we can prepare the initial state as $\rho = \prod_{k} \rho_k$, and the LOA of the total system is the product of the LOA for each $k$-mode, $\mathcal{G}(t)=\prod_{k}\mathcal{G}_k(t)= \prod_{k} r_k(t)e^{i\phi_k(t)}$, where $\phi_k(t)$ is the phase of the $k$-mode. By subtracting the dynamic phase, we can obtain the geometric phase, $\phi^g_k(t)=\phi_k(t)-\phi^d_k(t)$. For a model with specific symmetry, $\phi^g_k(t)$ are pinned to zero at $k$ = 0 and $\pi$. Then, DTOP can be defined as \cite{nu_Dprb,nu_Dprb1,nu_Dprb2,nu_Dprb3}
\begin{eqnarray}
\label{nu_D}
\nu_D (t)=\frac{1}{2\pi} \int_0^{\pi} \frac {\partial \phi_k^{g} (t)}{\partial k} \,dk.
\end{eqnarray}

The DTOP is a topological number induced by dynamical evolution \cite{nu_Dprb} and is considerably different from the ground-state topological number, i.e., its variation does not imply a changing bulk topology. Actually, the general quench approach does not influence the bulk topology at all \cite{prl115.236403,ncomm8336}.

\section{Generalized Loschmidt overlap amplitude.}
The generalizations of LOA are not unique. The previous results focus on the different aspects and give the different generalizations, which are applicable in some situations \cite{fidelity,metric,1d,2d,UMass1,UMass2}. One class of generalizations is based on quasi-distance approaches. They treat GLOA as a quasi-distance between states \cite{fidelity,metric}; for instance, the fidelity between $\rho(t)$ and $\rho(0)$, $\trace\sqrt{\sqrt{\rho(0)}\rho(t)\sqrt{\rho(0)}}$. GLOA based on this approach can be used in open system, however, such approaches lose the topological information encoded in the argument even in the unitary evolution process for pure states. The non-analytics of the rate function appearing at zero temperature become smooth at finite temperature, which suggests DQPT disappears at finite temperature in such generalizations \cite{fidelity,metric,SI}. Another possible class of generalizations are interferometric approaches. A common interferometric GLOA is defined as $\trace[\rho(0)U(t)]$ in unitary evolution for mixed states \cite{metric,1d,2d,SI}. DQPT can be observed at finite temperature based on this approach, however, to the best of our knowledge, the interferometric GLOA in nonunitary evolution has not been reported yet. The reason might be that to directly extend the above definition to an open system directly is ambiguous because of the lack of $U(t)$ in an open system. Due to the research carried out for a mixed-state geometric phase \cite{gphase,Vedralgphase,Vedralgphase2}, we can give a new type of GLOA for a mixed state in an open system. Here, we will use a kinematic approach for the geometric phase \cite{gphase}. The quantum trajectory of the arbitrary nondegenerate evolution with $N$ dimensional Hilbert space is
\begin{eqnarray}
\mathcal{P} : t \in [0,\tau] \rightarrow \rho(t) = \sum_{j=1}^{N} p_j(t)\ket{\phi_j(t)}\bra{\phi_j(t)},\label{traj}
\end{eqnarray}
where $p_j(t)$ and $\ket{\phi_j(t)}$ are the eigenvalues and eigenvectors of the density matrix, respectively.
Then, we define the GLOA as
\begin{eqnarray}
G(t) &=& \sum_{j=1}^{N} \sqrt{p_j(t)p_j(0)} \braket{\phi_j(0)|\phi_j(t)}
\nonumber \\
&&\times e^{-\int_{0}^{t}\braket{\phi_j(\tau)|\dot{\phi}_j(\tau)}\,d\tau}.
\end{eqnarray}
The related rate function is $g(t)=-\frac{1}{L}\ln |{G}(t)|^2$. We can find that the GLOA defined here can be reduced to a normal LOA modified dynamical phase in a pure state unitary evolution.
Importantly, the interesting physical quantities remain unchanged, i.e., we can still define them self-consistently.
For instance, the Fisher zero points and rate function do not change \cite{Fisherandrate}. Meanwhile, $G$ is the product of $G_k$, and DTOP can also be constructed via Eq.(\ref{nu_D}) when the system is noninteractiing and translationally invariant \cite{DTOPfromG}.

\section{Finite temperature effect.}
Usual theoretical analysis for DQPT requires that the initial state is the ground state of the Hamiltonian, which is in general difficult to realize in experimental processes. Generally, the initial state is a thermal state with low temperature. We will investigate the finite temperature effect on DQPT in this part via GLOA. As an explicit example, we consider the two-banded model with translation invariance \cite{nu_Dprb}. The Hamiltonian is $H= \sum_{k} \hat{H}_k=\vec{h}_k\cdot\vec{\sigma}$, where $\vec{h}_k$ is the three-component vector of momentum $k$ and $\vec{\sigma}$ is the vector of the Pauli matrix. The eigenvalue of the Hamiltonian is $\pm\epsilon_k=\pm|\vec{h}_k|$. We prepare the initial state as the thermal state with an inverse temperature $\beta$ for the Hamiltonian $\hat{H}_k^{i}=\vec{h}_k^{i}\cdot\vec{\sigma}$,
\begin{eqnarray}
\label{rho}
\rho_k = \frac{e^{\beta\epsilon_k^{i}}}{2\cosh\beta\epsilon_k^{i}}\ket{i_k^{-}}\bra{i_k^{-}}+\frac{e^{-\beta\epsilon_k^{i}}}{2\cosh\beta\epsilon_k^{i}}\ket{i_k^{+}}\bra{i_k^{+}},
\end{eqnarray}
where $\ket{i_k^{-}}$ and $\ket{i_k^{+}}$ are the lower and higher bands before quenching, respectively. Then, we apply a sudden quench to the system. The Hamiltonian after quenching is $\hat{H}_k^{f}=\vec{h}_k^{f}\cdot\vec{\sigma}$. Then, $G_k(t)$ can be computed explicitly,

\begin{widetext}
\begin{eqnarray}
G_k(t) &=&\frac{e^{\beta\epsilon_k^{i}}}{2\cosh\beta\epsilon_k^{i}}(|g_k|^{2}e^{i\epsilon_k^{f}t}
+ |e_k|^{2}e^{-i\epsilon_k^{f}t})
 e^{i\epsilon_k^{f}(|e_k|^{2}-|g_k|^{2})t} +
\frac{e^{-\beta\epsilon_k^{i}}}{2\cosh\beta\epsilon_k^{i}}(|e_k|^{2}e^{i\epsilon_k^{f}t}
+ |g_k|^{2}e^{-i\epsilon_k^{f}t})e^{i\epsilon_k^{f}(|g_k|^{2}-|e_k|^{2})t},
\nonumber \\
\end{eqnarray}
\end{widetext}
where $|e_k|^2=\frac{1}{2}(1-\hat{h}_k^i\cdot\hat{h}_k^f)$, $|g_k|^2=\frac{1}{2}(1+\hat{h}_k^i\cdot\hat{h}_k^f)$ and $\hat{h}_k^i=\vec{h}_k^i/|\vec{h}_k^i|$, $\hat{h}_k^f=\vec{h}_k^f/|\vec{h}_k^f|$.

As a benchmark, we consider the 1D transverse field Ising model (TFIM),
\begin{eqnarray}
H=-\frac{J}{2} \sum_{i=1}^{L}\sigma_i^{x}\sigma_{i+1}^{x}+\frac{h}{2}\sum_{i=1}^{L}\sigma_i^{z},
\end{eqnarray}
where $L$ is the number of sites, $J$ represents the coupling between the nearest spins and $h$ is the transverse field. For convenience, we consider the periodic boundary condition $\sigma_1=\sigma_{L+1}$ and set $J = 1$. It can be converted equivalently into a two-banded fermionic model with $\vec{h}_k=(0,\sin k,\cos k-h)$ via the Jordan-Wigner transform\cite{McCoy}. TFIM is in a ferromagnetic phase when $|h| < 1$ and in a paramagnetic phase when $|h| > 1$ \cite{McCoy}. Strictly speaking, $\rho_k$ in Eq. (\ref{rho}) is a four-dimensional matrix spanned by $\ket{00}_{k,-k}$,$\ket{11}_{k,-k}$, $\ket{01}_{k,-k}$ and $\ket{10}_{k,-k}$, and the Hamiltonian matrix is given by
\begin{eqnarray}
\hat{H}_k=
\left[
\begin{matrix}
  \vec{h}_k\cdot\vec{\sigma} & 0_{2\times 2} \\
  0_{2 \times 2} & 0_{2\times 2}
\end{matrix}
\right],
\end{eqnarray}
where $0_{2\times 2}$ denotes a two-dimensional zero matrix. The block structure of the Hamiltonian matrix is due to parity conservation and we can only consider the even occupation subspace, which is a two-dimensional matrix spanned by $\ket{00}_{k,-k}$ and $\ket{11}_{k,-k}$.

It is easy to verify that the phase $\phi^g_k$ is still pinned to zero at $k = 0$ and $\pi$. This leads to a periodic structure of $\phi^g_k$, which means DTOP can also be constructed as Eq.(\ref{nu_D}) at arbitrary temperature and remains well-quantized.

DQPT occurs when there is at least a critical $k_c$ that satisfies $|e_k|^{2}=|g_k|^{2}$ , with the corresponding critical time $t_{c,n}=(2n-1)\pi/2\epsilon_{k_c}^{f},n\in\mathbb{N}^*$. The geometric phase at momentum $k_c$ jumps $\pi$ at the critical time. This implies that the change in DTOP is the sufficient and necessary condition of DQPT. Furthermore, the change in DTOP at the critical time is only determined by the sign of $\beta$ and the sign of the slope $s_{k_c}=\partial_k |e_k|^2|_{k_c}$,
\begin{eqnarray}
\Delta\nu_D(t_c)&=&\lim_{\tau\to0^{+}}[\nu_D(t_c+\tau)-\nu_D(t_c-\tau)]
\nonumber \\
&=&\sign(s_{k_c})\sign(\beta).
\end{eqnarray}
This relation is very similar to Eq. (10) in Ref. [\onlinecite{nu_Dprb}]. The proof is straightforward and identical to the proof of Eq. (10) in Ref. [\onlinecite{nu_Dprb}].

\begin{figure}
\flushleft
\subfigure[]{\includegraphics[width=0.23\textwidth]{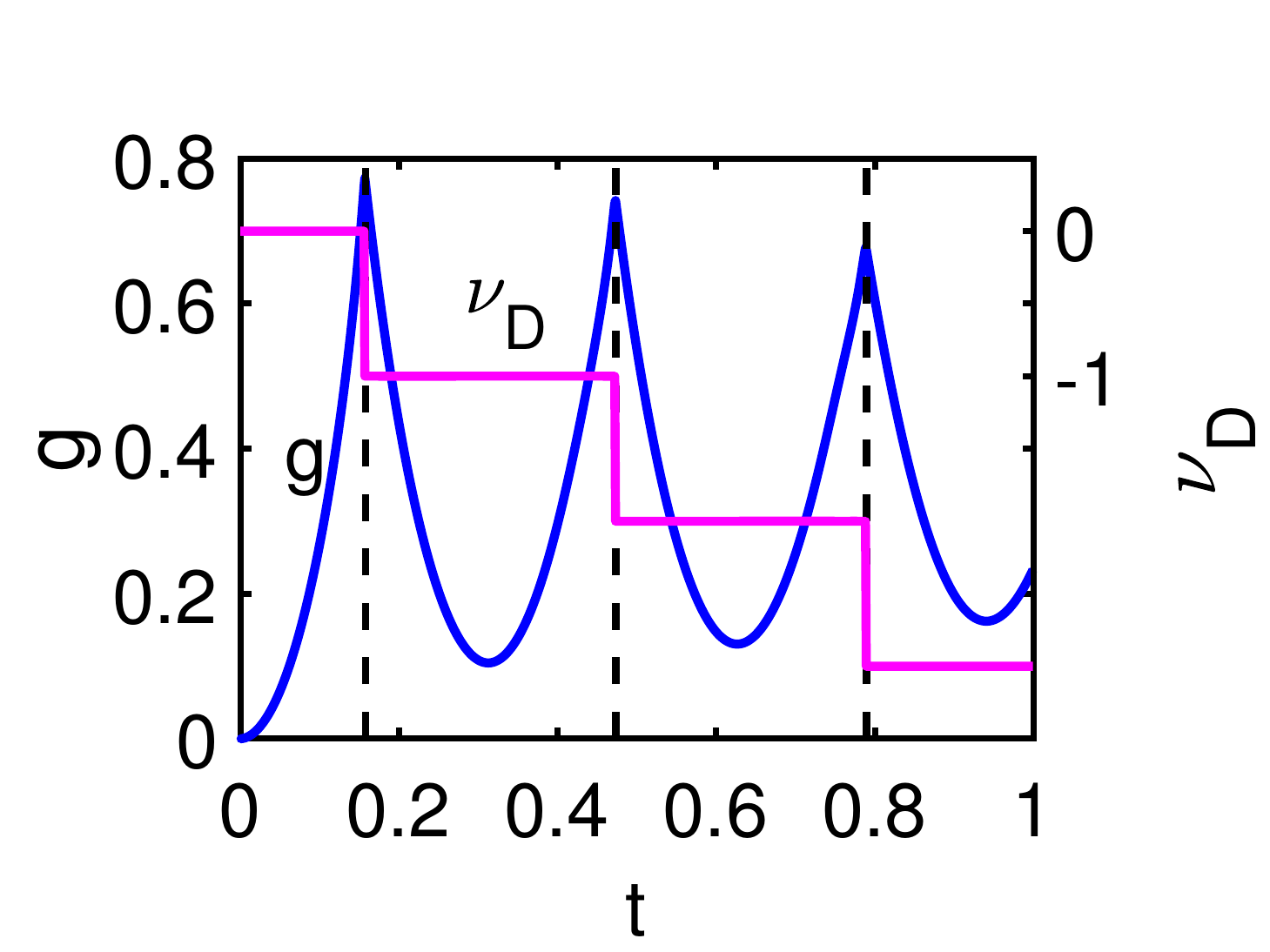}}
\subfigure[]{\includegraphics[width=0.23\textwidth]{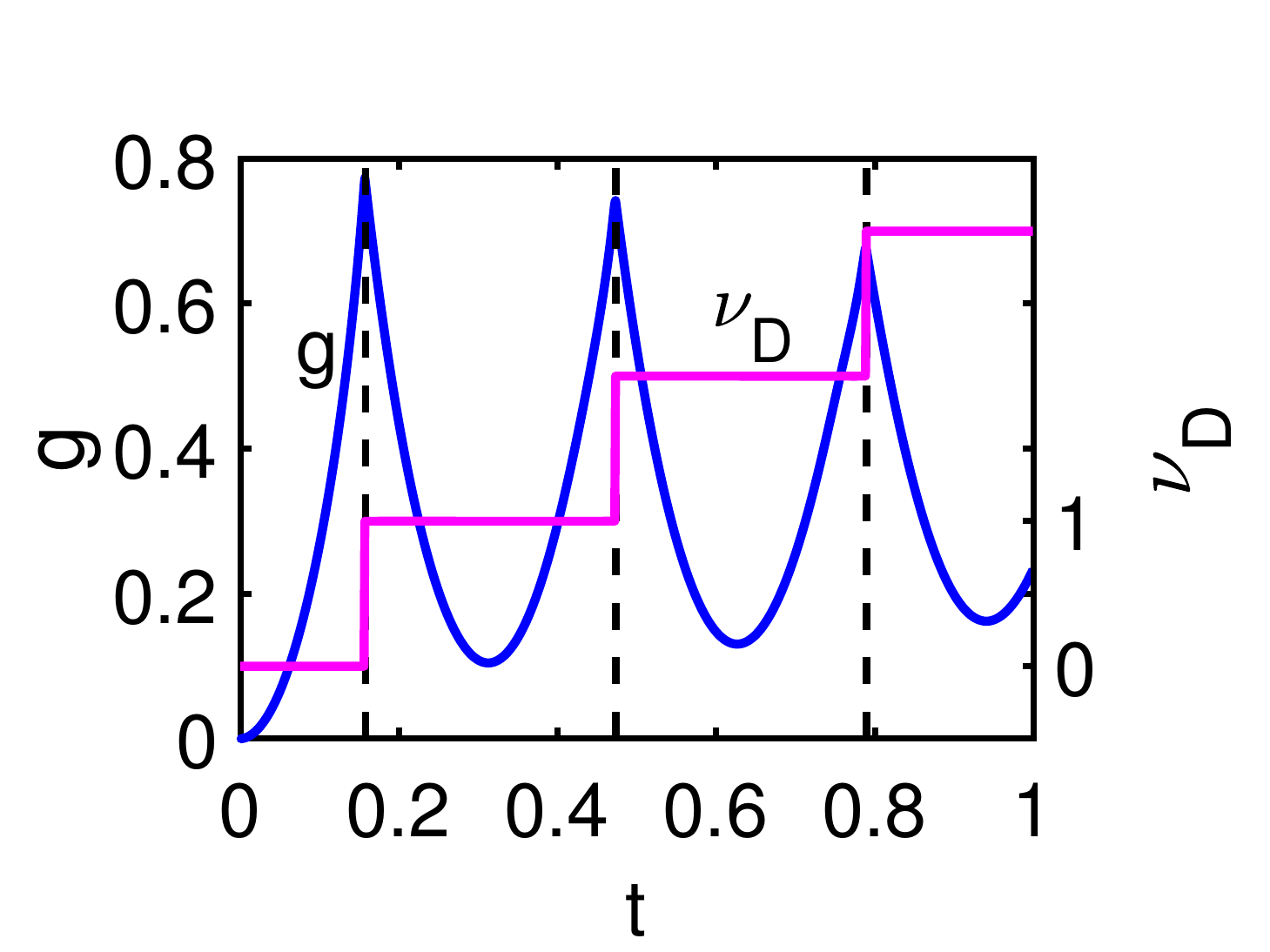}}
\caption{Time dependence of DTOP and rate function with $h_i = 0$, $h_f = 10$, (a) $\beta =1$ and (b) $\beta = -1$. Blue Lines are rate functions and pink lines are DTOPs. Black dash lines cross the $x$-axis at $t_{c,n}=(2n-1)\pi/2\epsilon_{k_c}^{f},n\in\mathbb{N}^*$.}\label{figure1}
\end{figure}

We set $h_i = 0$ and $h_f = 10$ with an inverse temperature of 1 and -1. We can expect a DQPT to occur since we quench across different phases. The time dependence of the DTOP and rate function are shown in Fig.\ref{figure1}. We can observe the non-analyticity of the rate function, coinciding with the changing DTOP. Moreover, the DTOP changes its value depending on the sign of the temperature.

At the end of this section, we give a short remark on the comparison of our GLOA and GLOA in Refs. [\onlinecite{1d,2d}]. All GLOAs take a very similar form but show a considerable difference for the unitary evolution for mixed states. The GLOA proposed here is equivalent to finding the $\mathcal{U}(t)$ that satisfies the parallel transport condition \cite{gphase,Vedralgphase,Vedralgphase2}
\begin{eqnarray}
\bra{\phi_j(t)}\dot{\mathcal{U}}(t)\mathcal{U}^\dag(t)\ket{\phi_j(t)}, j = 1,2,...N\label{PTC}
\end{eqnarray}
where the notations are identical to those in Eq.(\ref{traj}) to eliminate the free-phase factor of the density matrix \cite{free}. We can also express our GLOA as $\trace[\rho(0)\mathcal{U}(t)]$. In this sense, two proposals for GLOA take the same form but with a different evolution operator. Both evolution operators make the density matrix evolve in the same way. Our GLOA chooses an evolution operator that satisfies the parallel transport condition. Strictly speaking, the geometric phase defined in Ref. [\onlinecite{1d}] is $U(1)$ gauge invariant for the state of a specific purification. The geometric phase here is equivalent to an average of the geometric phases for each different pure state trajectory in a reasonable way \cite{SDiehl}.

\section{Natural dissipation effect.}
In addition to the finite temperature effect, one important effect in experiments is dissipation, i.e.\ the system is not closed and has interactions with the environment. Here, we consider the situation where the Born-Markov approximation works. Then, the evolution in an open system satisfies quantum master equation\cite{Revdiss},
\begin{eqnarray}
\dot{\rho}=-i[H,\rho]+\sum_{\mu}L_\mu \rho L_\mu^{\dag} - \frac{1}{2}\{L_\mu^{\dag}L_\mu,\rho\},
\end{eqnarray}
where $H$ is the Hamiltonian of the system describing coherent dynamics, and $L_\mu$ is a Lindblad operator describing damping dynamics. The spectrum of the quantum master equation is seminegative, indicating that the density matrix will be damped to a steady state, which is absent in a closed system \cite{Revdiss}.

It is difficult to solve TFIM with natural dissipation such as $L_\mu=\sigma_z^\mu$, where $\sigma_z^\mu$ is the pauli matrix of site $\mu$ because the master equation is not quadratic in the fermionic model after Jordan-Wigner transformation. For simplicity, we use GLOA to analyze the robustness of singularity under decoherence in the two-banded fermion model, which is TFIM after Jordan-Wigner transformation with fermion leakage dissipation $L_\mu=\sqrt{\gamma_\minus} a_\mu$ and injection dissipation $L_\mu=\sqrt{\gamma_\plus} a_\mu^{\dag}$, where $a_\mu^{\dag}$ and $a_\mu$ are the annihilation and creation operators for the fermion at $\mu $-th site \cite{fidelity}.

In the model described above, transverse-invariance symmetry does not break and different $k$-modes are still separable; however, parity symmetry is broken due to dissipation, which means we must consider the full dynamics for the four-dimensional density matrix. We consider the following quench process. The initial state is still prepared as Eq. (\ref{rho}), and then a sudden quench is applied. The system is governed by the following quantum master equation after quenching,
\begin{eqnarray}
\dot{\rho}_{k}&=&-i[\hat{H}_k,\rho_{k}] \nonumber\\
&&+\gamma_\minus \sum_{\sigma=\pm} a_{\sigma k} \rho_{k}  a_{\sigma k}^{\dag} - \frac{1}{2}\{ a_{\sigma k}^{\dag} a_{\sigma k},\rho_{k} \}
\nonumber\\
&&+\gamma_\plus \sum_{\sigma=\pm}a_{\sigma k}^{\dag} \rho_{k} a_{\sigma k} - \frac{1}{2}\{ a_{\sigma k}  a_{\sigma k}^{\dag},\rho_{k} \}.
\end{eqnarray}

\begin{figure}
\flushleft
\subfigure[]{\includegraphics[width=0.23\textwidth]{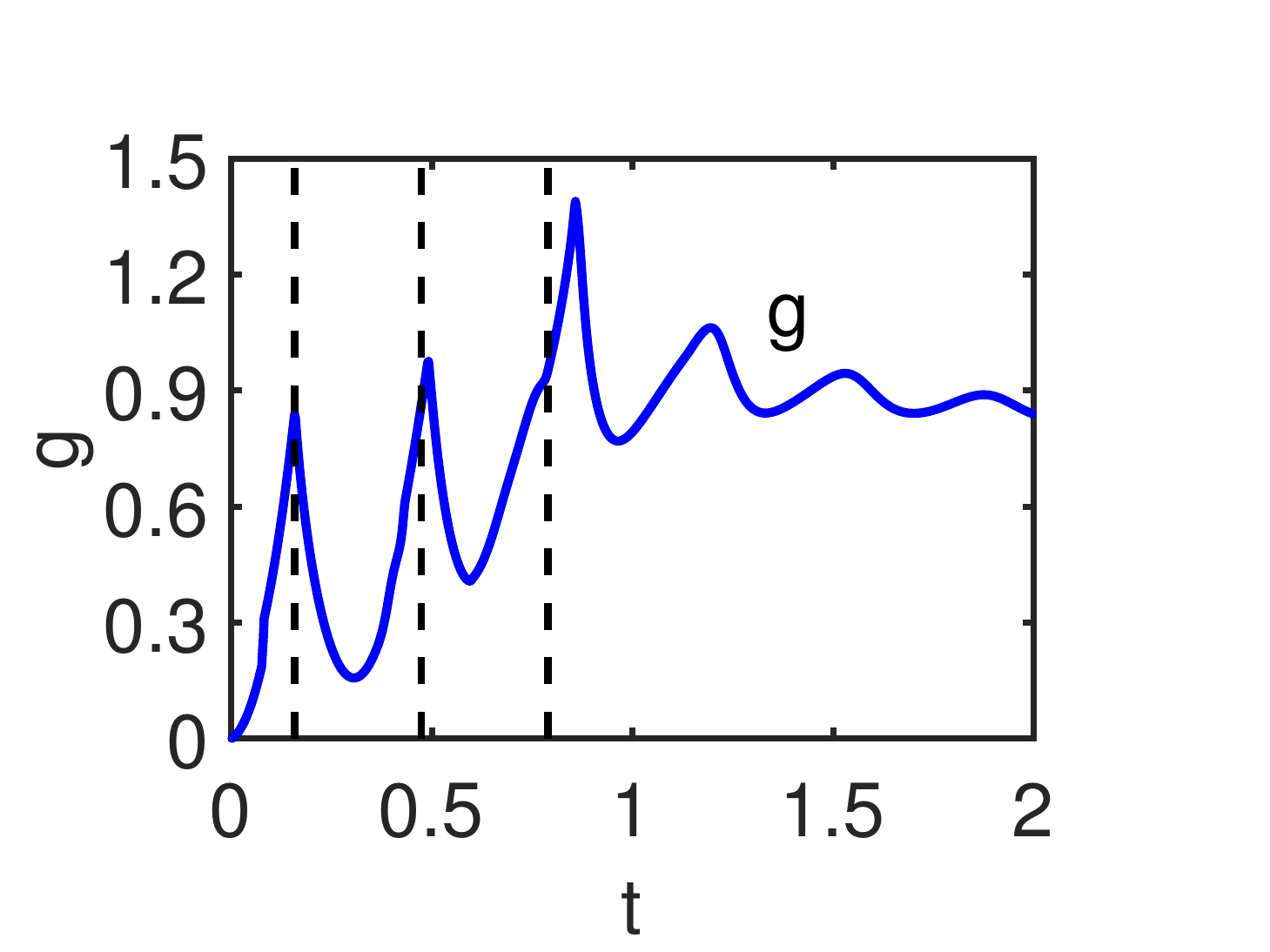}}
\subfigure[]{\includegraphics[width=0.23\textwidth]{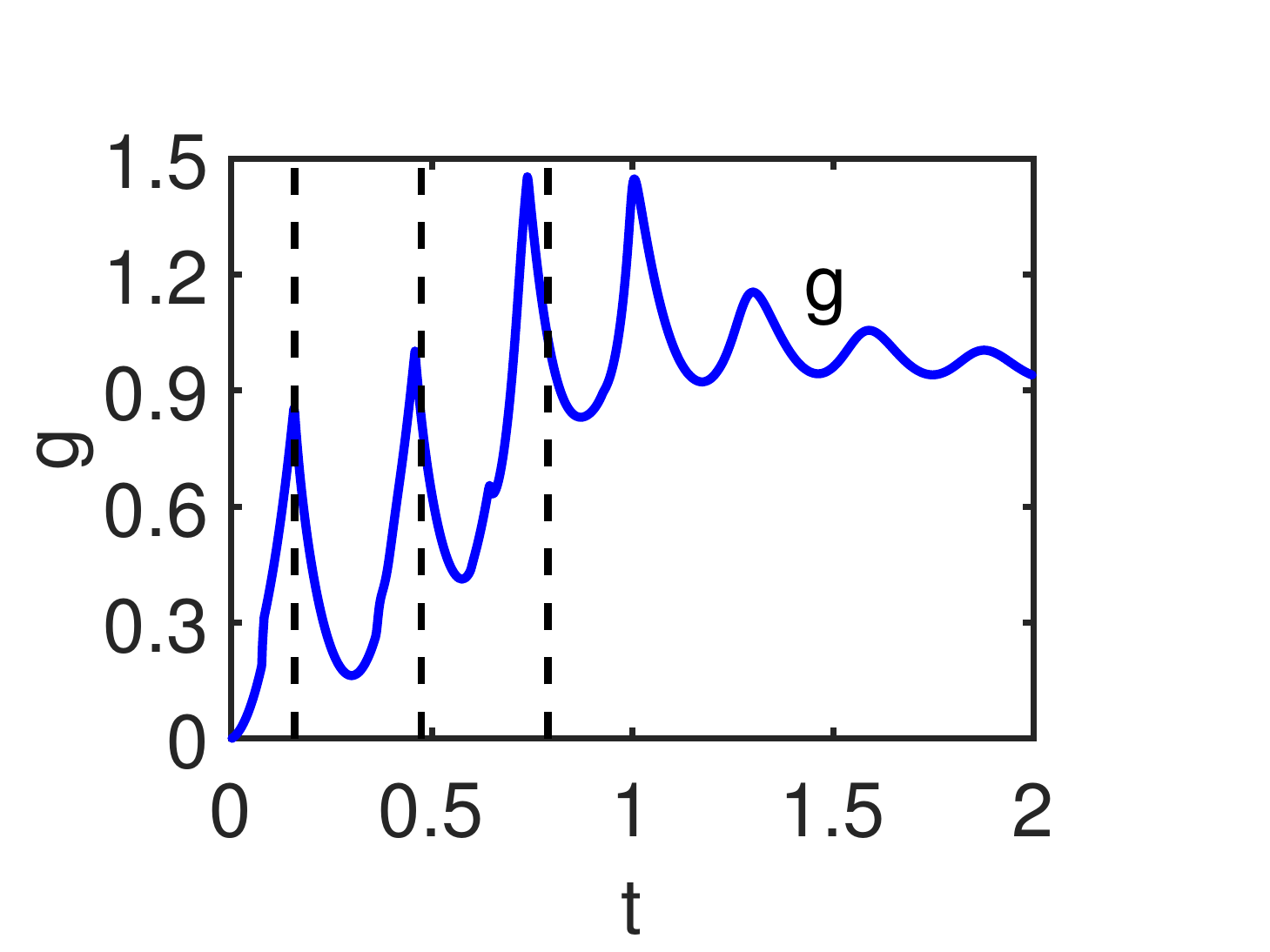}}
\subfigure[]{\includegraphics[width=0.23\textwidth]{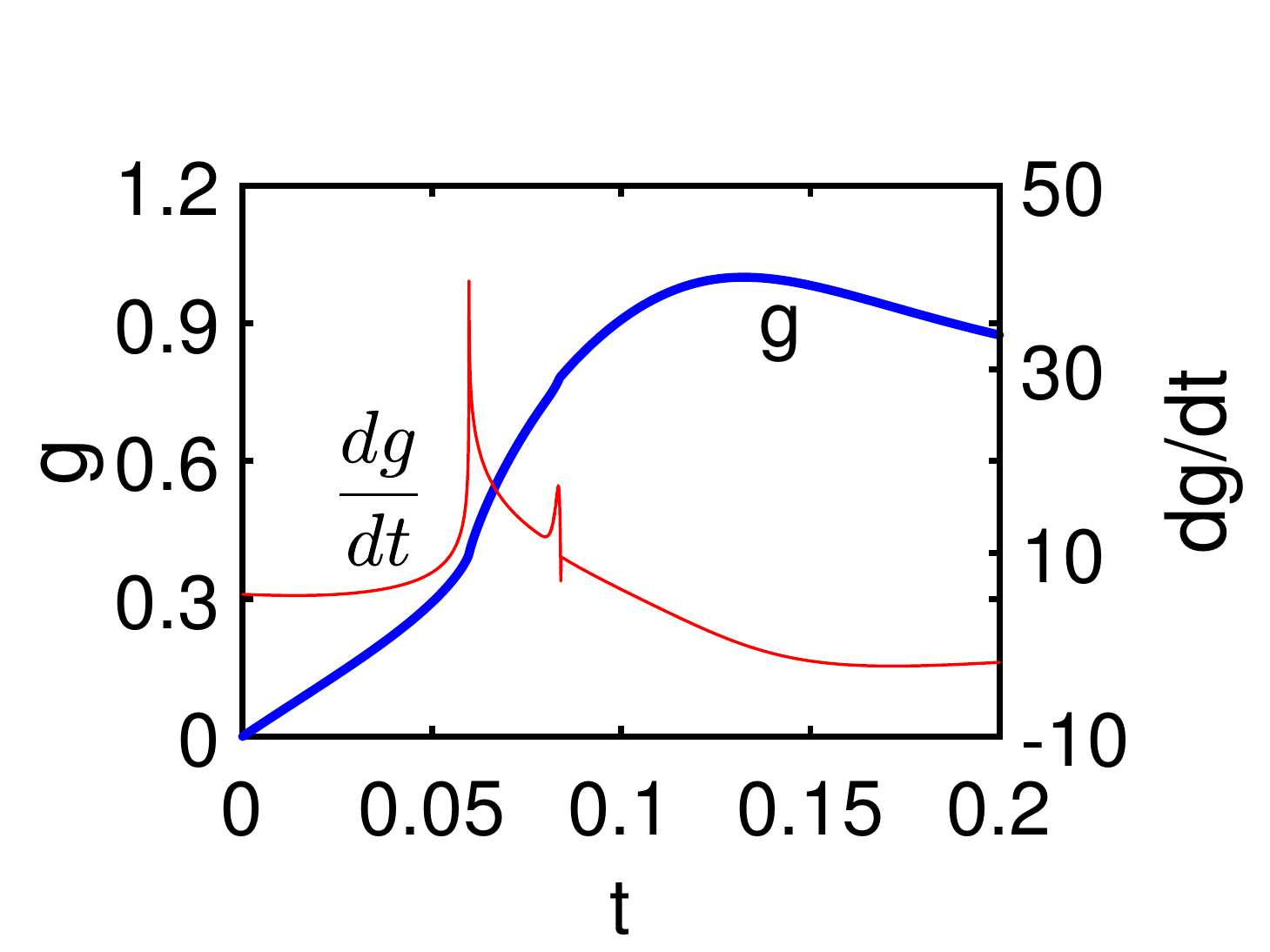}}
\subfigure[]{\includegraphics[width=0.23\textwidth]{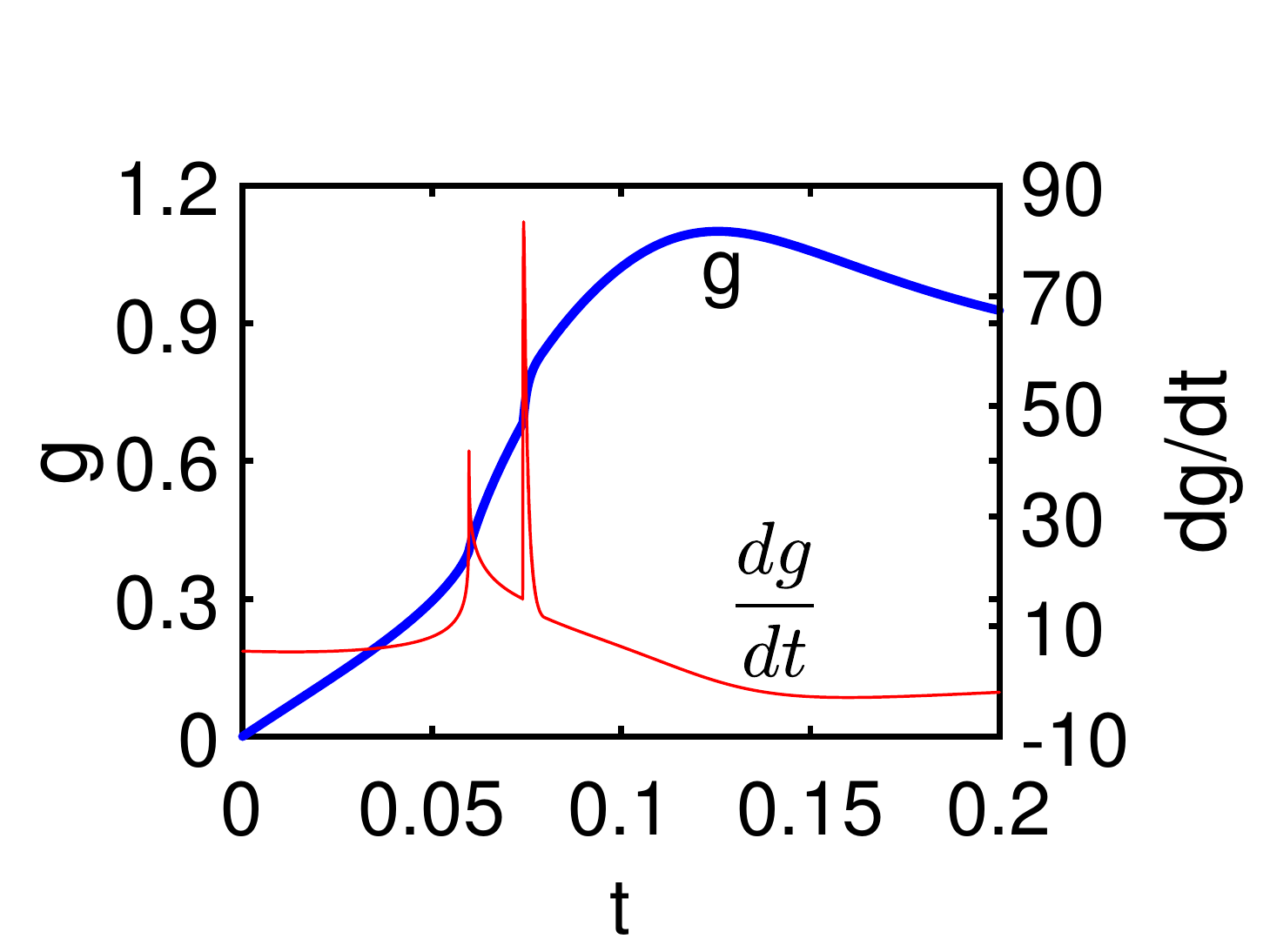}}
\caption{Time dependence of the rate function with quenching across different phases, $\beta=1$,$h_i = 0$ and $h_f = 10$,(a) $\gamma_{\plus}=0.1$, $\gamma_{\minus}=1$, (b) $\gamma_{\plus}=1$, $\gamma_{\minus}=0.1$, (c) $\gamma_{\plus}=1$, $\gamma_{\minus}=10$, (d) $\gamma_{\plus}=10$, $\gamma_{\minus}=1$. Blue lines denote the rate functions, and the red lines denote the derivatives of the rate functions. Black dash lines in (a) and (b) denote $t_{c,n}=(2n-1)\pi/2\epsilon_{k_c}^{f},n\in\mathbb{N}^*$, which is the critical time without dissipation.} \label{naturaldiff}
\end{figure}

\begin{figure}
\flushleft
\subfigure[]{\includegraphics[width=0.23\textwidth]{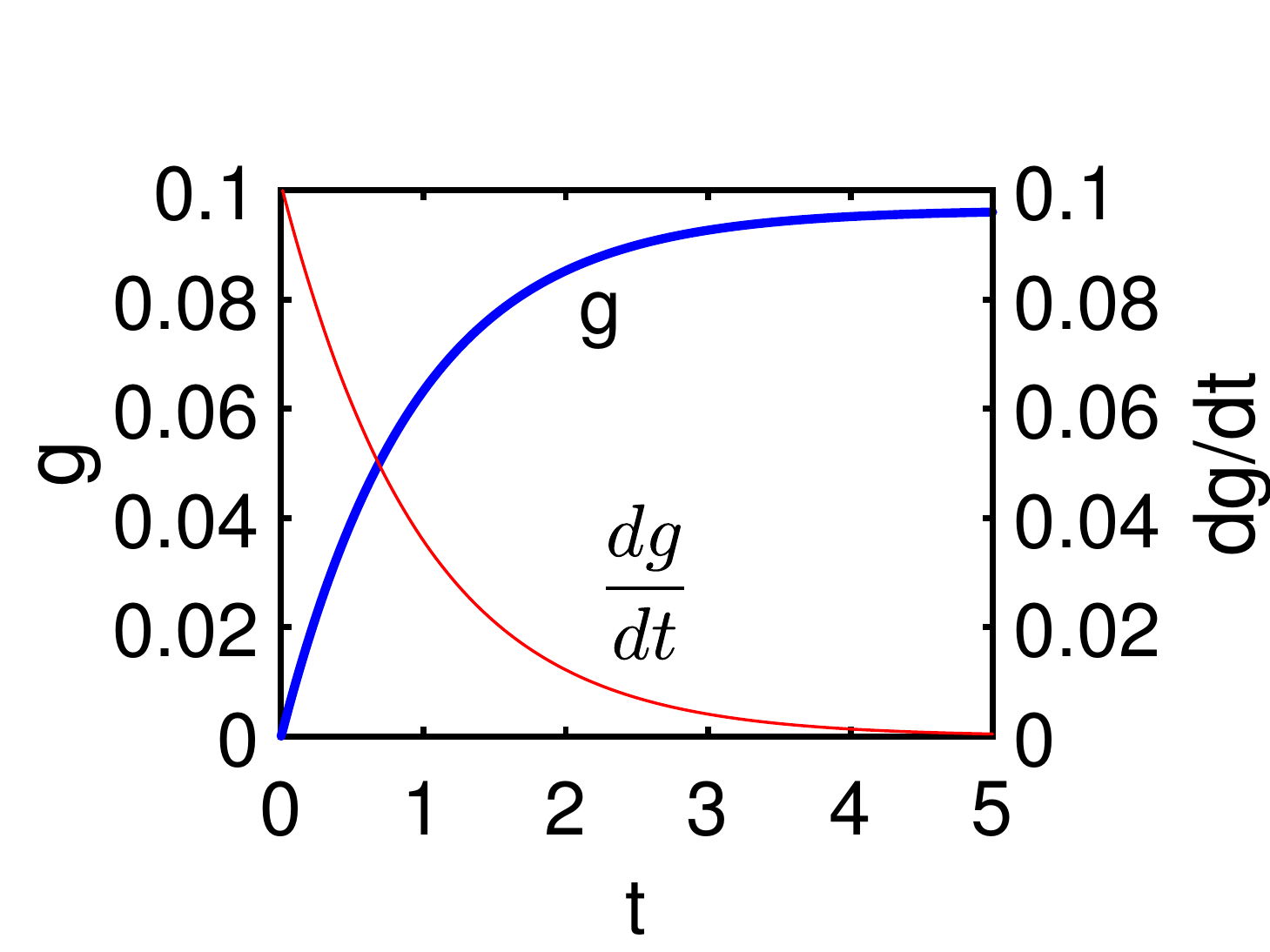}}
\subfigure[]{\includegraphics[width=0.23\textwidth]{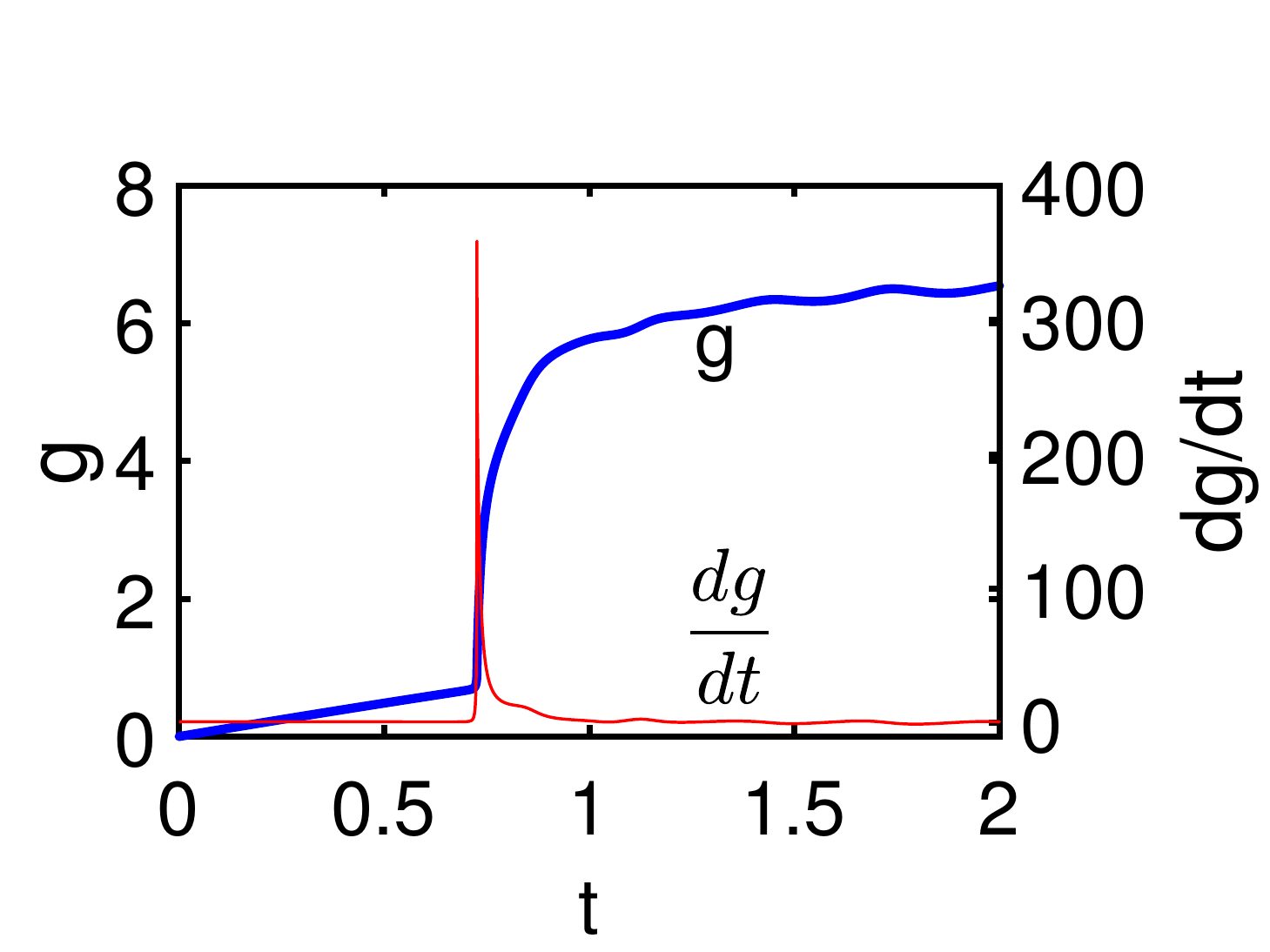}}
\subfigure[]{\includegraphics[width=0.23\textwidth]{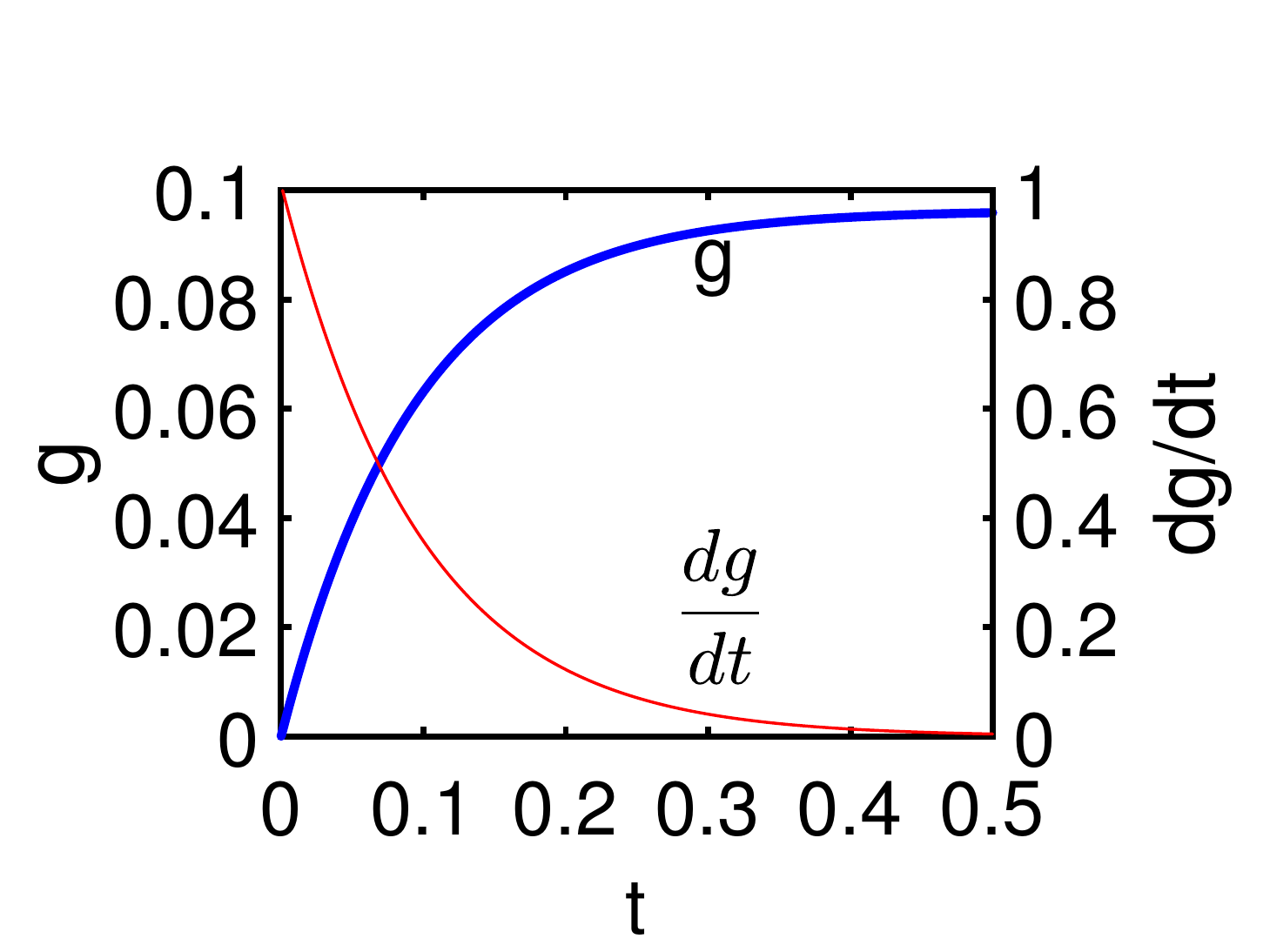}}
\subfigure[]{\includegraphics[width=0.23\textwidth]{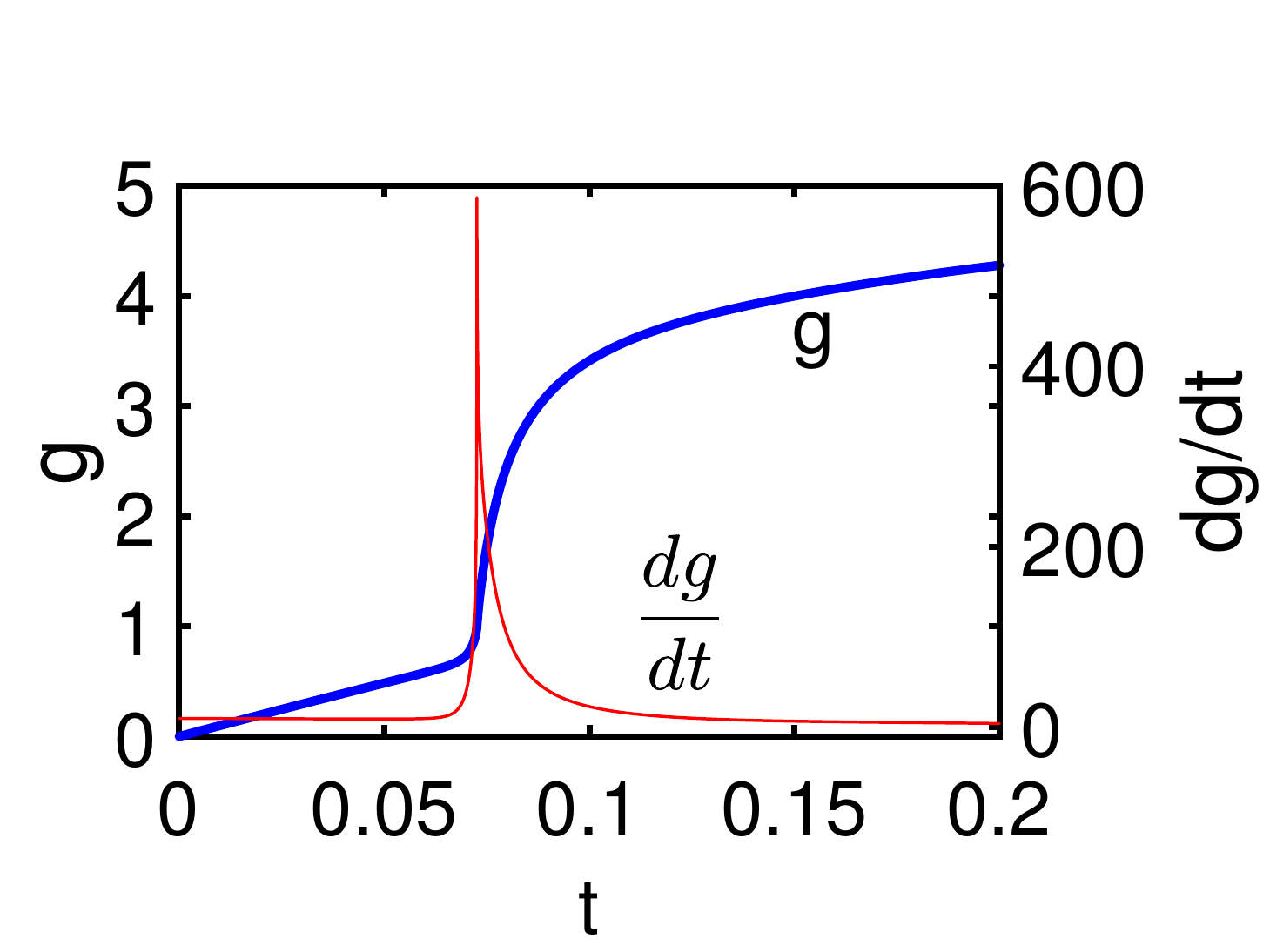}}
\caption{Time dependence of the rate function with quenching in the same phase, $\beta=1$,$h_i = 10$ and $h_f = 10$,(a) $\gamma_{\plus}=0.1$, $\gamma_{\minus}=1$, (b) $\gamma_{\plus}=1$, $\gamma_{\minus}=0.1$, (c) $\gamma_{\plus}=1$, $\gamma_{\minus}=10$, (d) $\gamma_{\plus}=10$, $\gamma_{\minus}=1$. Blue lines denote rate functions, and red lines denote the derivatives of the rate functions.}\label{naturalsame}
\end{figure}

We show the numerical results for the rate functions and their derivatives \cite{SI} when we quench across different phases in Fig. \ref{naturaldiff} and the quench in the same phase in Fig. \ref{naturalsame}. We can observe that DQPT persists at a high temperature and weak enough dissipation up to several period, however, with a shifted critical time, as shown in Fig. 2(a) and Fig. 2(b). In addition, a new kind of DQPT emerges without the influence of coherent dynamics, suggesting that it is a dissipation-induced DQPT, corresponding to the non-analytic points of the derivatives in Fig. 2(c, d) and Fig. 3(b, d).

\section{Engineered driven-dissipation.}
On the one hand, unavoidable natural dissipation leads to system decoherence. On the other hand, some exciting nonequilibrium models can be realized in an AMO system via the controlled driven-dissipation, which is unachievable in condensed matter \cite{Revdiss}. Here, we consider the 1D spinless dissipation topological superconductor model with translation invariance. It is a long-time mean field approximation result of a novel realistic model \cite{NJP,diss2d,ZollerNatph}. The master equation is quadratic in the approximation and we can solve it analytically. We assume the Hamiltonian of the system is zero, i.e., the dynamics are only determined by Lindblad operators. In momentum space, the Lindblad operator takes the form $L_k=v_ka_{k}^{\dag}-u_ka_{-k}$ and the density matrix of the steady state is $\rho=\prod_{k}\rho_k$ where $\rho_k$ is a four-dimensional matrix, as mentioned above. Topological information is encoded in the even occupation part, which is proportional to $\frac{1}{2}(I_{2\times2}+\vec{n}_k\cdot\vec{\sigma})$. If all $\vec{n}_k$ are nonzero and the system has chiral symmetry, the topological index can be defined as \cite{NJP,diss2d,ZollerNatph}
\begin{eqnarray}
\nu =\frac{1}{2\pi} \int_{-\pi}^{\pi} \vec{a}\cdot({\hat{n}}_k\times \frac {\partial {\hat{n}}_k}{\partial k}) \,dk\in \mathbf{Z},
\end{eqnarray}
where $\hat{n}_k=\vec{n}_k/|\vec{n}_k|$ is the pseudo spin and $\vec{a}$ satisfies $\vec{a}\cdot\vec{n}_k = 0$ for all $k$.

A topological phase transition can occur in open systems, surprisingly, with much richer phenomena than closed systems. It can be realized via the closure of the dissipative gap, purity gap or both; for more discussions see Ref. [\onlinecite{NJP}]. We establish here the relation between the DQPT and nonequilibrium topological phase transition with three explicit benchmarks, with the potential to become the new paradigm for phase transitions in open systems.

\begin{widetext}

\begin{figure}[htpb]
\flushleft
\subfigure[]{\includegraphics[width=0.33\textwidth]{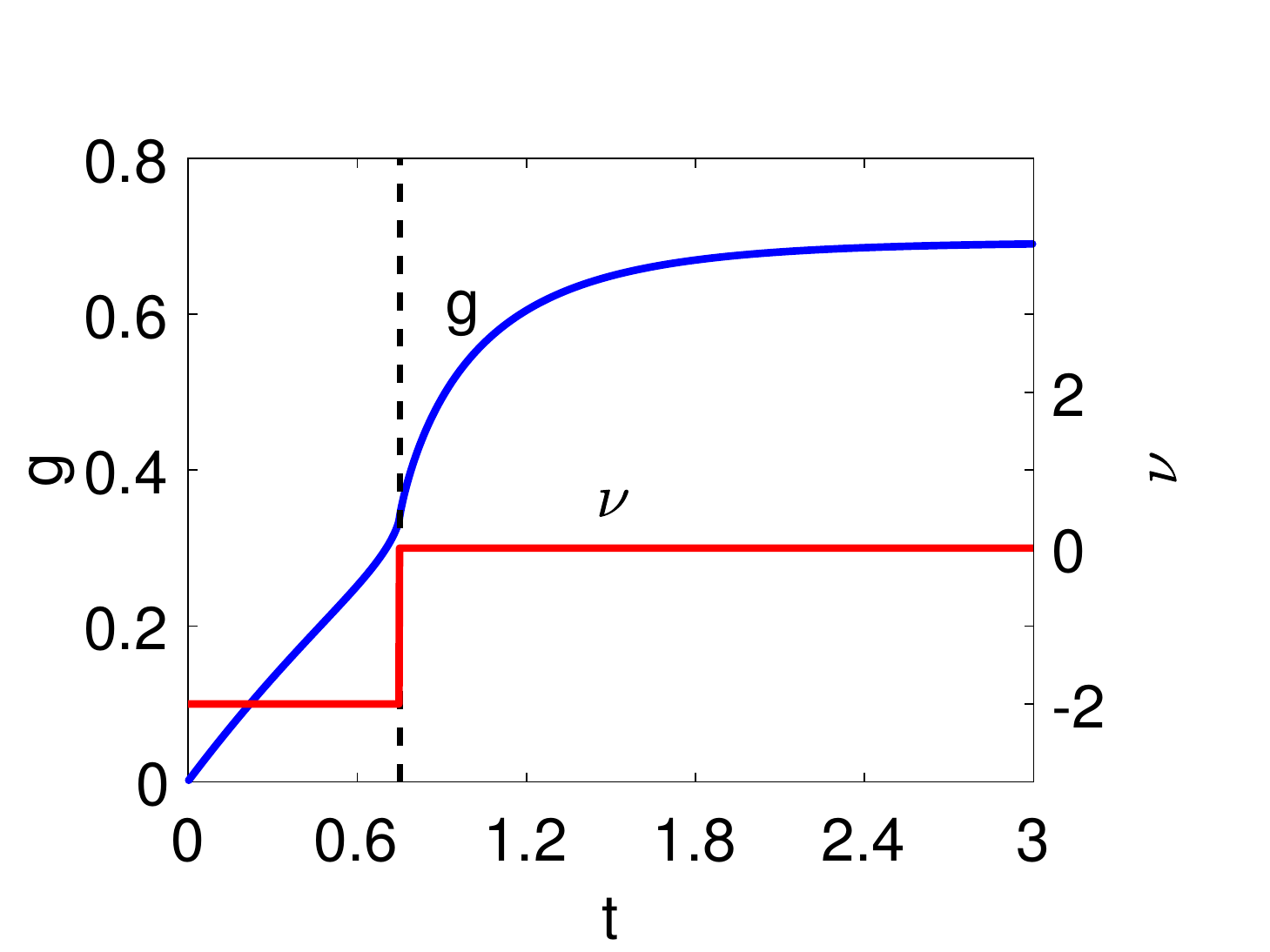}}
\subfigure[]{\includegraphics[width=0.33\textwidth]{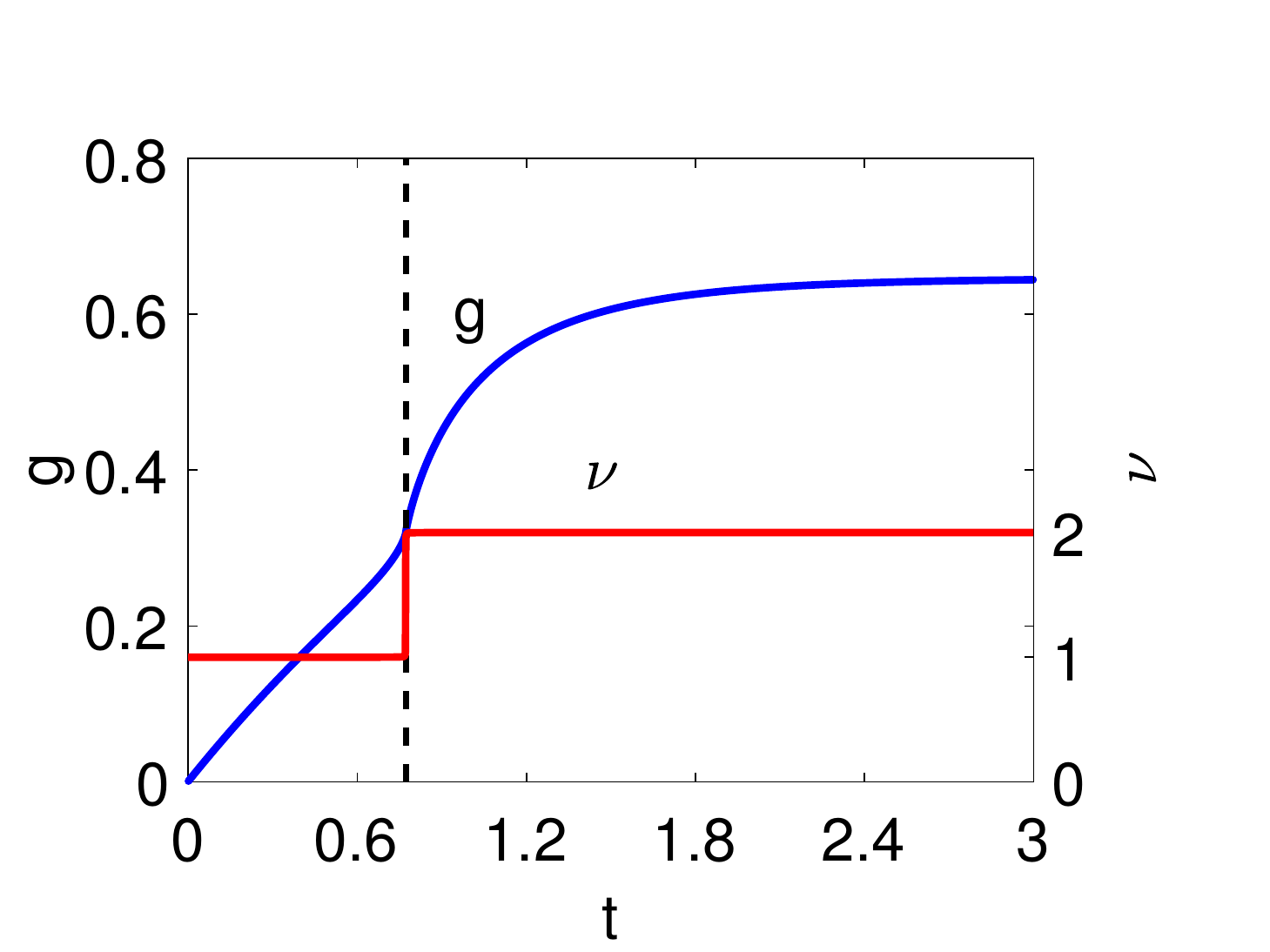}}
\subfigure[]{\includegraphics[width=0.33\textwidth]{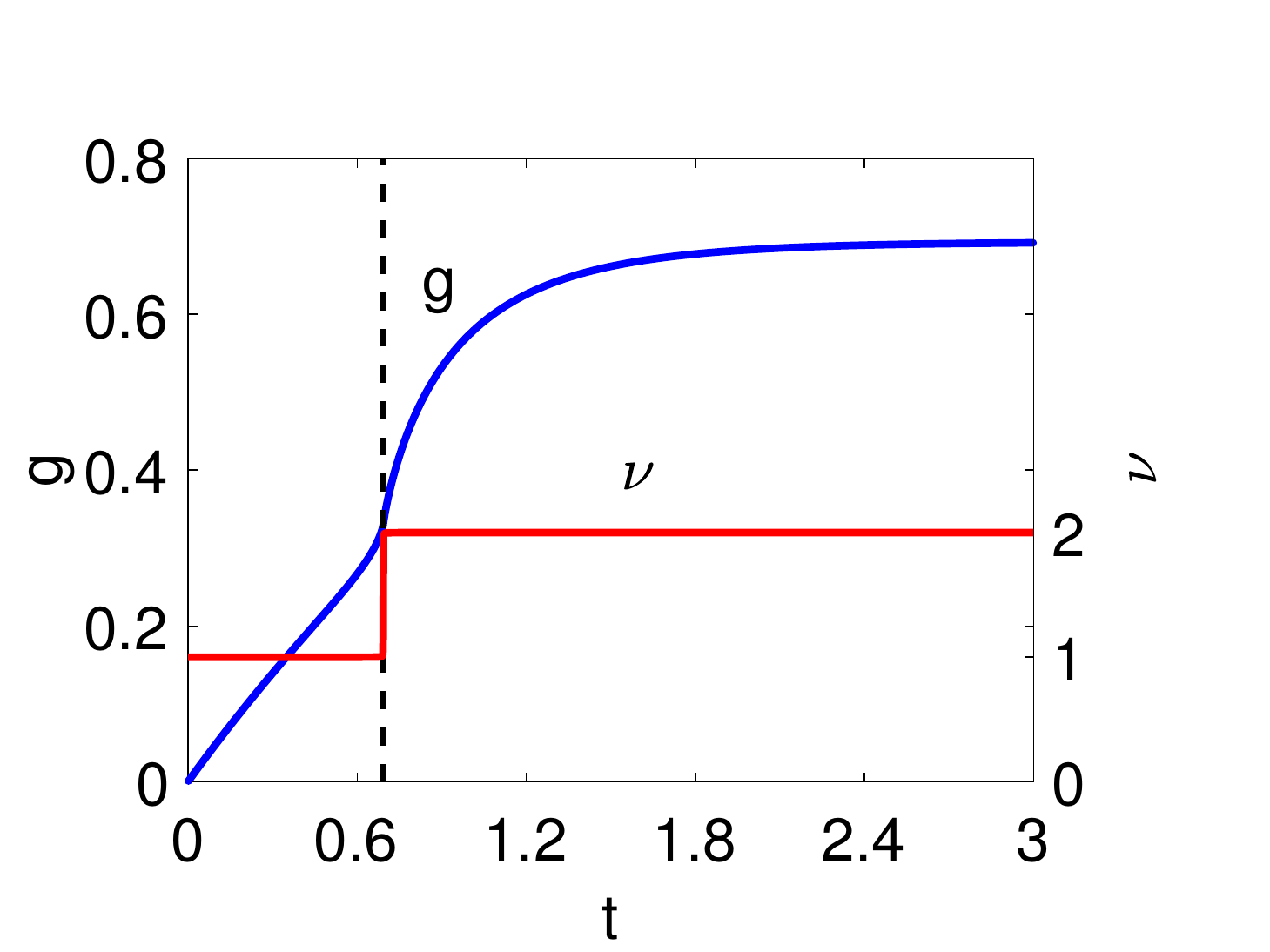}}
\caption{Time dependence of bulk-topology index and rate function with $\kappa_i = 0$, $\kappa_f = 10$ for three driven-dissipation benchmarks. Blue lines denote rate functions and red lines denote bulk-topological numbers. Black dash lines are normal to the $x$-axis and cross $\nu$ and rate functions at $t_c$ in all three benchmarks.} \label{engineer}
\end{figure}

\end{widetext}

Next, we present three benchmarks with quenching across different phases, which can enable one to identify the nonequilibrium phase transitions described by three kinds of gap closure methods \cite{SI,NJP}. Fig.\ref{engineer} displays the time dependence of $\nu$ and the rate function. A DQPT occurs in all these benchmarks, suggesting that a DQPT remains a dynamical probe for a nonequilibrium phase transition. However, there are some dramatic differences compared to the unitary model. First, the non-analyticity is not revived due to the loss of coherent dynamics. Second, DTOP is not an order parameter since the $x$ component of the pseudo spin is always zero in the models and the DTOP never changes. Third, the bulk topological property can change in the dissipation models if we quench across the different topological regions, which is forbidden in a unitary quench process.

We can understand the changing bulk topology in two ways. Macroscopically, the initial state and long-time steady state have different topological properties in dissipation models, so the bulk topological number must change at some time. If we focus on the microscopic details of this difference, we can obtain information that is much more useful. When quench dynamics are unitary, $\hat{n}_k(t)$ and $\vec{n}_k(t)$ evolve continuously, with their length conserved. The direct result of such properties is that the changes in the bulk topology property are forbidden \cite{prl115.236403,ncomm8336}. However, in the dissipation model, the length of $\vec{n}_k(t)$ is not conversed and appears in the denominator of the expression for $\hat{n}_k(t)$. At critical time $t_c$, there exists at least one $k_c$ to make $\vec{n}_{k_c}(t_c)$ vanish, which results in an ill-defined $\hat{n}_{k_c}(t_c)$. This provides the possibility for a discontinuous change in the winding number. Furthermore, the critical time where the rate function is non-analytic is identical to the bulk topological properties changing time (see Fig.\ref{engineer}), which implies that they share the same reason, i.e., the ill-defined pseudo spin \cite{SI1}.

\section{Conclusion and discussion.}
In summary, we extend the conception of LOA to the most general quantum evolution process, with the GLOA proposed by us still containing the argument information. We compute GLOA in three explicit examples. We observe that a DQPT persists at an arbitrary finite temperature and a weak enough decoherence. We also find a new type of DQPT in dissipation models. Moreover, the GLOA has a large variety of potential applications.
First, it is possible to carry out theoretical analysis for a DQPT in a system with other kinds of dissipation phase transitions; for instance, dissipation-induced superradiant phase transitions appearing in a Rabi model\cite{MBPlenioPRA}. Investigation of the GLOA in such systems can enable one to gain clearer information about the role of dissipation in phase transitions and a DQPT. Another possible interesting topic is the classical correspondence of GLOA. Notice that such correspondence is very similar between a partition function and the Loschmidt overlap amplitude for pure state unitary evolution \cite{MHeyl}. By identifying the thermodynamic phase transition counterparts to a DQPT in a nonunitary phase transition one can obtain deep insights for the GLOA and DQPT.
Furthermore, the GLOA we have proposed is experimentally accessible in systems such as fermionic ultracold atoms in an optical lattice \cite{natpexpDQPT,tomography}. The technique of reconstructing the time-evolving many-body state provides full information for the GLOA.

\section{Acknowledgements} This work was supported by the National Key R \& D Plan of China (Grant Nos. 2016YFA0302104,
2016YFA0300600), the National Natural Science Foundation of China (Grant Nos. 91536108, 11774406),
and Strategic Priority Research Program of Chinese Academy of
Sciences (Grant No. XDB28000000).


\begin{thebibliography}{9}
\bibitem{Fisherandrate}The only difference between GLOA and LOA is the phase factor when the state is pure and the dynamics are unitary. The rate function is only determined by the modular of LOA, which remains unchanged. Meanwhile, Fisher zero points which are the zero points of LOA also remain unchanged.
\bibitem{DTOPfromG}From the definition, it is easy to obtain ${G}(t)=\prod_{k}{G}_k(t)= \prod_{k} r_k(t)e^{i\phi^g_k(t)}$. Then, we can construct DTOP via Eq.(\ref{nu_D}).
\bibitem{free}When Hamiltonian is given, the trajectory of density matrix is determined. However, for each $j$, $\ket{\phi_j(t)}$ in Eq.(\ref{traj}) has a free phase factor undetermined.
\end{thebibliography}
\end{document}